# New Equation For Describing Time Dependence of Moon's Orbit Radius

Mikrajuddin Abdullah

Department of Physics, Bandung Institute of Technology

Jalan Ganesa 10 Bandung 40132, Indonesia

MIBE S&T Institute

Jalan Sembrani 19 Bandung, Indonesia

**Abstract**

I derived an equation to describe the dependence of Moon's orbit radius around the Earth. I obtained the radius changes with time according to a scalling equation, $R_{m-e} \propto t^{2/5}$. Using the equation I have been able to predict accurately some quantities that have been well accepted.

## I. INTORUDUCTION

The Moon orbits the Earth at an elliptic orbit with the apogee of 405,400 km and the perigee of 362,600 km. The average radius of the Moon's orbit is 384,399 km [1]. It is believed that the Moon has been created more that 30 Ma after the start of the Solar System [2]. Susprisingly, new findings have shown that the Moon's orbit radius is getting larger at a rate of about 37.8 mm/year [3,4]. The interesting question then, is there any formula to describe the change of Moon's orbit radius with time? How is the rate of change of the orbit radius at the past and the future?

The purpose of this short work is to derive an equation that describing the dependence of the Moon's orbit radius with time. Very surprising that the derived model is able to predict different quantities that well accepted at the present.

## II. MODELLING

Since the Earth contains fluid component (ocean and atmosphere), the fraction of which might be easily displaced by Moon or Sun gravitation, the shape of the Earth, especially on its surface, will sligthly vibrate continuously (assumed to be random). This change will disturbate slightly the strength of the Earth's gravitational force on the Moon to imply the distance of the Moon's orbit radius will change slightly. The rate of the Moon's orbit radius change will depend on the instantaneous Moon's orbit radius. As the Moon's orbit gets large, the efffect of Earth mass variation of the gravitational force experienced by the Moon decreases so that the rate of Moon's orbit change will decrease too.

I simplify the model by assuming that the dominant gravitational attraction experienced by the Moon originates only from the Earth. Gravitational disturbances that might originate from other bodies such as the Sun or asteroids are neglected. We also assumed that the fluctuation in the Earth gravitational field affects only the orbital motion of the Moon and does not affect the Moon's rotational orbit.

For modeling purpose, I propose the following assumptions. I assume the total mass of the Earth, $M_e$, can be divided into two components: the fixed mass $M_e - \delta$ and the displaceable mass $\delta$. The fixed mass has a constant distance from the Moon, $R_{m-e}$ (Moon's orbit radius) and the displaceable mass has variable distance from the Moon, $R_{m-e} + \Delta r$ with $\Delta r$ is a random variable, $-r_e \leq \Delta r \leq r_e$, and $r_e$ is the Earth's radius. The instantaneous potential energy of the Moon due to Earth gravitation becomes

$$U_m = -\frac{GM_m(M_e - \delta)}{R_{m-e}} - \frac{GM_m \delta}{R_{m-e} + \Delta r}$$

$$= -\frac{GM_m M_e}{R_{m-e}} + \frac{GM_m \delta}{R_{m-e}}\left(1 - \frac{1}{1 + \Delta r / R_{m-e}}\right)$$

$$\approx -\frac{GM_m M_e}{R_{m-e}} + \frac{GM_m \delta}{R_{m-e}}\left(\frac{\Delta r}{R_{m-e}} - \left(\frac{\Delta r}{R_{m-e}}\right)^2\right) \quad (1)$$

Thue the average potential energy of the Moon becomes

$$\langle U_m \rangle \approx -\frac{GM_m M_e}{R_{m-e}} + \frac{GM_m \delta}{R_{m-e}^2}\langle \Delta r \rangle - \frac{GM_m \delta}{R_{m-e}^3}\langle \Delta r \rangle^2 \quad (2)$$

Sincre $\Delta r$ is random then $\langle \Delta r \rangle = 0$ and Eq. (2) becomes

$$\langle U_m \rangle \approx -\frac{GM_m M_e}{R_{m-e}} - \frac{GM_m \delta}{R_{m-e}^3}\langle \Delta r \rangle^2 \quad (3)$$

Therefore, the presence of randomly Earth's mass displacement resulted in reduction in the Moon potential energy as much as

$$\Delta \langle U_m \rangle \approx -\frac{GM_m \delta}{R_{m-e}^3} \langle \Delta r \rangle^2 \qquad (4)$$

This reduction in potential energy can be compared to reduction in potential energy due to fluctuation of charge distribution of atom or molecules as described bythe van der Walls force. In term of atom where the attraction force is the coulomb force, the reduction in the potential energy satisfies $\Delta U \propto -1/R^6$ [5]. This potential energy is the sum of potential energy by slightly separated positive and negative charges from two atoms. For the gravitational force, since there are no gravitational charges, the change in the potential energy varies according to inverse of qubical distance.

Due to change in the potential energy of gravitation, there is a change in the gravitational force experiences by the Moon, satisfying

$$\Delta F = -\frac{d\Delta \langle U_m \rangle}{dR_{m-e}}$$

$$\approx \frac{3GM_m \delta}{R_{m-e}^4} \langle \Delta r \rangle^2 \qquad (5)$$

It is clear that the fluctuated gravitational force decreases if the Moon's orbit radius increases, resulting in decrease in the Earth-Moon force. Using the second Newton law of motion, the gravitational force fluctuation will generate fluctuation in the radial acceleration of the Moon as

$$M_m \frac{d^2 R_{m-e}}{dt^2} = \Delta F$$

$$M_m \frac{d^2 R_{m-e}}{dt^2} = \frac{3GM_m \delta}{R_{m-e}^4} \langle \Delta r \rangle^2$$

$$R_{m-e}^4 \frac{d^2 R_{m-e}}{dt^2} = 3G\delta \langle \Delta r \rangle^2 \qquad (6)$$

To obtain the solution of Eq. (6), let us temporarily assume

$$R_{m-e} = At^\beta \qquad (7)$$

so that

$$\frac{d^2 R_{m-e}}{dt^2} = \beta(\beta-1)At^{\beta-2}$$

Substituting into Eq. (6) one has

$$A^5 \beta(\beta-1)t^{4\beta+\beta-2} = 3G\delta\langle\Delta r\rangle^2$$

Equalizing power and factor in both sides we have

$$5\beta - 2 = 0$$

$$A^5 \beta(\beta-1) = 3G\delta\langle\Delta r\rangle^2$$

The first equation resulting

$$\beta = \frac{2}{5}$$

and the secons equation resulting

$$A = 1.66\left(G\delta\langle\Delta r\rangle^2\right)^{1/5}$$

Therefore we obtain the time dependece of Moon's orbit radius as

$$R_{m-e}(t) = 1.66\left(G\delta\langle\Delta r\rangle^2\right)^{1/5} t^{2/5} \qquad (8)$$

### III. DISCUSSION

The question is, how to justify that Eq. (8) correctly descibes the change of Moon's orbit radius with time. We have a data that at present the Moon's orbit radius increases with a rate of around 35 mm/year = $3.5\times 10^{-5}$ km/year (some report says 38 mm/year). The rate change of Moon's orbit radius is obtained by differentiating of Eq. (8) as

$$\frac{dR_{m-e}(t)}{dt} = 0.663\left(G\delta\langle\Delta r\rangle^2\right)^{1/5} t^{-3/5} \qquad (9)$$

We assume that the Moon start to revolve the Earth since its created for about 4.51 billion years ago[6]. Starting from that time, we assume the Moon's orbit radius started to increase with time. Using the data $dR_{m-e}/dt \approx 3.5 \times 10^{-5}$ km/year and $t \approx 4.51\times 10^9$ years we find

$$3.5\times 10^{-5} = 0.663\left(G\delta\langle\Delta r\rangle^2\right)^{1/5}\left(4.51\times 10^9\right)^{-3/5} \text{ to result}$$

$$\left(G\delta\langle\Delta r\rangle^2\right)^{1/5} = 32.74 \qquad (10)$$

Then Eq. (8) can be written as

$$R_{m-e}(t) = 54.35 t^{2/5} \qquad (11)$$

Again, using the present age of the Moon, we estimete the present Moon's orbit radius as $R_{m-e}(t) = 54.37 \times (4.51 \times 10^9)^{2/5}$ = 395,220 km. Suprisingly, this value is excatly similar the to present average Moon's orbit radius of 384,399 km [1] as a prove that Eq. (8) or (11) might be correctly describe the time dependence of the Moon's orbit radius.

We can use the Eq. (10) to estimate the mass of the ocean displaced by the Moon or the Sun gravitation. We use $G$ = 6.67 × $10^{-11}$ m$^3$/kg s$^2$ = 6.64× $10^{-7}$km$^3$/kg year$^2$. We assumed the displacement of sea is approximatedwith the heigh of sea tide, and here we use the highest sea tide in Nova Scotia of around 16.2 m so that $\langle \Delta r^2 \rangle \approx 0.00026$ km$^2$ [1]. Using Eq. (10) we get the estimated $\delta \approx 2.7 \times 10^{17}$ kg. The total mass of water in the ocean is around 1.347 × $10^{21}$ kg [7,8]. Therefore, the displaced water mass accounts for approximately 0.02% of the total mass of the ocean. In my opinion, this figure is very acceptable. The average tidal height in open sea is around 0.6 m [1]. Thus, the approximated displaced wated by the tide is around $\Delta M_e \approx (4\pi r_e^2 \times 0.6) \times 1,000$ = 3.1 × $10^{17}$ kg. Very surprising that this value is nearly exactly similar to our estimation for $\delta$.

Although the present model looks simple, however, we have been able to predict accurately the well accepted data, especially the present Moon's orbit radius and the mass of water displaced by tide.

**CONCLUSION**

The proposed equation has been able to exactly estimated different quantities that have been well accepted by scientists. Interestingly, there have been no free parameters used to estimate such quantities, instead all paremeter are the measured ones, to indicate that the proposed equation is in the right track.